# Capillary Flow Printing of Submicron Carbon Nanotube Transistors

Brittany N. Smith[a,§], Faris M. Albarghouthi[a,§], James L. Doherty[a], Xuancheng Pei[a], Quentin Macfarlane[a], Matthew Salfity[a], Daniel Badia[a], Marc Pascual[c], Pascal Boncenne[c], Nathan Bigan[c], Amin M'Barki[c], and Aaron D. Franklin[a,b]

[a] Electrical and Computer Engineering Department, Duke University, Durham, NC 27708, USA

[b] Chemistry Department, Duke University, Durham, NC 27708, USA

[c] Hummink Inc, 75003 Paris, France

[*] Correspondence to: aaron.franklin@duke.edu, TEL: +1-919-681-9471

[§] B.N.S. and F.M.A. contributed equally to this paper

# Abstract

Although printed transistors have a wide range of applications, the limited resolution of printing techniques (10-30 µm) has been a barrier to advancement and scaling, particularly down to submicron dimensions. While previous works have shown creative approaches to realizing submicron channel lengths with printing, reliance on chemical processes unique to specific inks or tedious post-processing limit their applicability. Here, we report the use of capillary flow printing (CFP) to repeatably create fully printed submicron carbon nanotube thin-film transistors (CNT-TFTs) without chemical modification or physical manipulation post-printing. The versatility of this printing technique is demonstrated by printing conducting, semiconducting, and insulating inks on several types of substrates ($SiO_2$, Kapton, and paper) and through the fabrication of various TFT device (contacting/gating) architectures. Notably, CFP of these CNT-TFTs yielded on-currents of 1.12 mA/mm when back gated on Si/$SiO_2$, and 490 µA/mm when side gated through ion gel on Kapton, demonstrating the strong transistor performance achievable with CFP. Mechanical bending and sweep rate resilience of devices printed on Kapton show the wide utility of CFP-fabricated devices for flexible applications. This work highlights the ability of CFP as a viable fabrication method for submicron electronics through cleanroom-free printing techniques.

# Main

The proliferation of additive manufacturing has transformed many products through the lower cost and more environmentally sustainable printing of 3D objects[1,2]. This has spurred pursuit of printing for the electronics industry, where there is increasing attention to the environmental impact of incumbent technologies[3]. Whether it is thin-film transistors (TFTs) for displays or nanoscale field-effect transistors (FETs) for microprocessors, transistor fabrication processes are leading contributors to fluorinated greenhouse gas emission and resultant climate change[4–8]. Meanwhile, printed transistors have been demonstrated using eco-friendly processing and materials that are completely recyclable[6,9]. It has long been a goal to realize fully printed transistors with performance and size that are competitive with commercial devices from silicon or metal-oxide semiconductors[10,11].

Motivation for realizing printed transistors is not necessarily for replacing sub-10 nm, high-performance technologies. There are opportunities in the TFT field for revolutionizing the display industry with printed backplane electronics[12–14], burgeoning applications for printed transistors in the expanding Internet of Things and biomedical wearables spaces[15], and the potential for printing devices into the back-end-of-line (BEOL) for monolithically integrated functionality in chips[16,17]. A promising assortment of printable semiconductor inks have been formulated: indium gallium zinc oxide (IGZO)[18–20], organics[21,22], 2D materials[23–25], and carbon nanotubes (CNTs)[26–28]. Among these candidates, CNT thin films offer high carrier mobility, chemical and mechanical robustness, and processability[29–31]. Yet, there have remained limits in achievable dimensional scaling and performance of fully printed transistors, from CNTs or any other semiconductor.

The smallest reliable resolution offered through current printing techniques is around 10-30 µm using aerosol jet printing (AJP) or inkjet printing (IJP) – far larger than that required to achieve submicron channel lengths[32]. This resolution constraint manifests itself in both the linewidth of a single printed feature, as well as the gap between two printed lines, which would determine the channel length in a printed transistor. In addition, while inkjet printing offers precise liquid delivery with small ink volume requirements, it includes challenges with droplet dynamics (wetting, merging, and satellite drops)[32,33]. Similarly, AJP enables wide ink compatibility, but overspray and resolution constraints present a significant hinderance to the feature sizes that can be achieved[34]. Besides these direct-write printing techniques, there are sheet or roll-to-roll approaches such as gravure printing[35,36] that provide exceptional throughput capabilities but are limited to typical resolutions of 50 μm, or 2-5 μm in the most aggressively scaled cases[37,38].

Several research groups have developed techniques for sidestepping the resolution constraints of printing technologies to achieve submicron dimensions. Approaches have included ink-to-ink repulsion (where ink solvent polarities cause two overlaying inks to repel from one another, yielding a micron-scale gap)[39], the use of self-assembled monolayers (which rely on the same repulsion technique to create micron-scale gaps)[40,41], post-print line-splitting[42], and dip-pen nanolithography (which is often a patterning lithography step requiring extensive chemical treatments and not a direct-write printing technique)[43,44]. While these work-arounds have enabled demonstrations of submicron gaps between conductive traces, there are challenges of repeatability due to process complexity, reliance on homogeneous chemical functionalization, and the difficulty of achieving fine control of transistor channel dimensions because of variability in line edge roughness. Thus, there remains a need for a printing technique that can directly print submicron channel lengths in a repeatable and simple manner.

In this work, we utilize a capillary flow printing (CFP) technique, derived from atomic force microscopy (AFM) technology, to demonstrate repeatable submicron printing of CNT transistors in a facile and scalable manner. This technique bypasses issues such as overspray, satellite droplets, and low throughput to enable the direct control of spacing between printed features at submicron levels. We harness the CFP technique to print a variety of inks (metallic, semiconducting, and insulating) to fabricate submicron CNT transistors. Specifically, the effects of various gating architectures, contact geometries, post-processing conditions, and substrates on device performance are studied, showcasing the versatility of this tool at creating submicron devices, including fully printed and flexible TFTs. Overall, the use of this capillary flow printing technology to fabricate, optimize, and characterize submicron transistors is a step toward the implementation of low-cost, rapid, and cleanroom-free development of scaled transistors.

**Capillary flow printing of nanomaterials**

The CFP technology (Hummink NAZCA printer) borrows techniques from atomic force microscopy wherein a macroresonator tuning fork detects deflections of a tip scratching across a surface at high frequencies. When a pipette (with a tip on the order of 100s of nm to a few microns) is placed on the end of the resonator and filled with a printable ink, the tip is able to effectively write or draw patterns of this ink onto a substrate, much like a fountain pen[45,46]. While there have been other uses of capillarity (i.e., capillary forces) to achieve printing of inks into 2D or 3D patterns[47,48], the actual printing was enabled using other forces, such as pressure or physical extrusion. In this work, a tabletop CFP system (Fig. 1a-c) was used to rapidly print nanomaterial patterns, resulting in fully printed TFTs with submicron channel lengths. When the pipette is in contact with the substrate surface, an ink meniscus is formed, meaning that when the pipette (or substrate) is moved, the meniscus leaves behind a "printed" pattern with virtually any ink (Fig. 1d). As the diameter of the chosen pipette tip and the "spreading factor" (a software-based parameter that accounts for the degree to which the ink spreads as a result of its viscosity) are changed, so too is the morphology of the printed line (line width dimensions, overlap between lines, etc.), enabling a fine degree of control over print features.

In this CFP tool, computer-aided design (CAD) files of the patterns (Fig. S1a-b) are printed through movement in the X-Y plane of the platen holding the substrate (Fig. S1c). To demonstrate the fine control and versatility of this tool, several patterns were printed using different inks. Within the context of transistor fabrication, silver nanoparticle (AgNP) source and drain line spacing, or channel length ($L_{ch}$), may be tuned down to the 100s of nanometers with uniformity throughout the spacing, as seen in Fig. 1e. To demonstrate the printer's fine control capabilities, two AgNP electrodes were printed with a ~340 nm gap between them, as measured by SEM (Fig. 1f), and shown to not be shorted together (i.e., electrically an open circuit) when tested. Importantly, this was achieved without any surface modification of the substrate or printed traces – the AgNP lines were simply printed directly onto the substrate with the 340 nm gap. Additionally, a Duke logo was printed from the AgNP ink, with a total pattern width of 200 μm, showcasing the ability for fine control of the substrate to create high-resolution printed patterns (Fig. 1g). This highlights a key attribute of this printing technology: the ability to create printed patterns at submicron, and even sub-500 nm scales, which is particularly useful for the miniaturization of printed transistors.

Capillary flow printing is compatible with inks having viscosities up to 100,000 cP, with no special accommodation needed from the user to enable material deposition with well-defined edges (i.e., no need for adjustment of piezoelectric driver or atomization current/flow parameter settings used in other printers). This aspect of the printer is shown in Figures 1h-i, where ion gel of the same formulation was deposited using CFP and AJP, highlighting an important distinction where CFP produces thinner traces in height and width as well as crisp line edges without overspray (i.e., errant aerosolized particles from AJP). Additionally, CFP enables the printing of SU-8 6000.5 without dilution or other treatments – a reduction in processing steps from AJP printing of SU-8 (Fig. S2a)[49]. To improve printability and reduce nozzle clogging in CFP, solvent volatility, surface tension, and nanoparticle dimensions must be accounted and optimized for when printing custom inks. When these factors are adjusted for, the application space for CFP-printed devices is wide. For example, scaling down dimensions enables the printing of many devices in a small area (132 devices were printed in 1 mm$^2$) as well as the shrinking of biosensors to reduce the amount of liquid needed for conformal coverage of all devices (Fig. 1j, S2b-c).

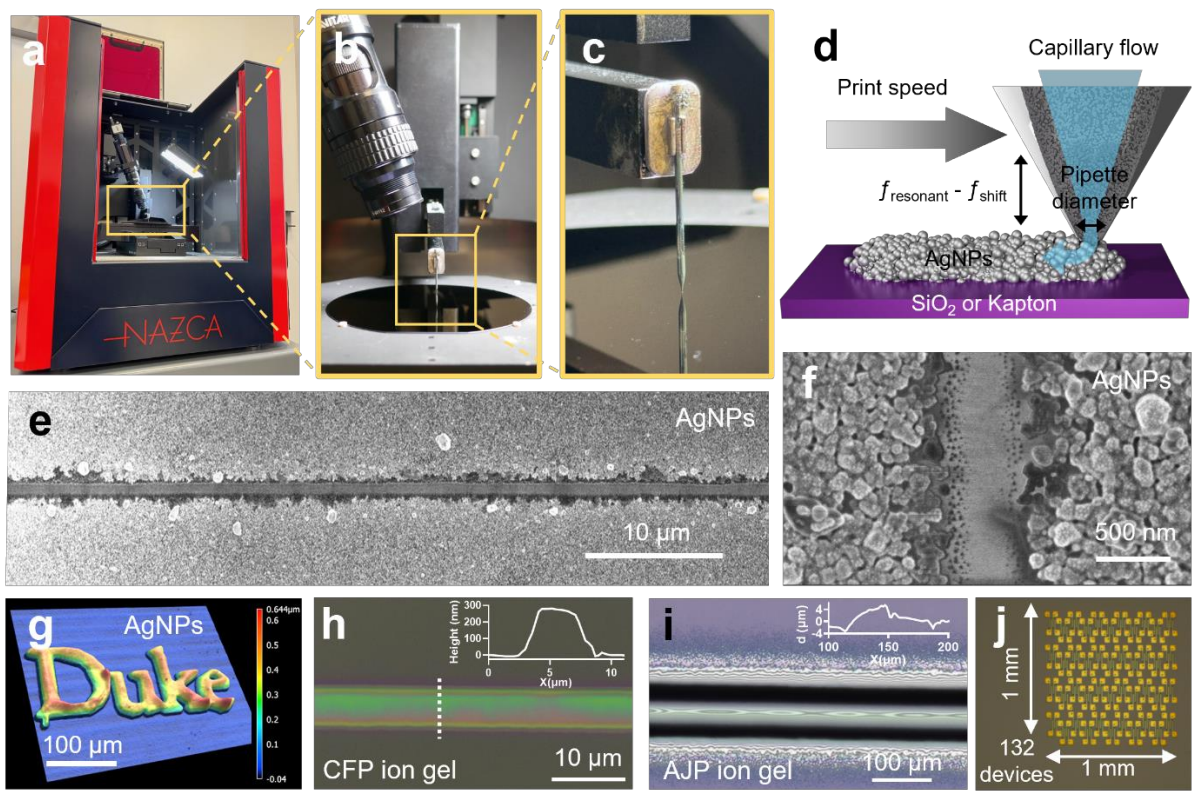

**Figure 1. Capillary flow printer capabilities.** Pictures of (a) Hummink NAZCA capillary flow printer, (b) pipette printing on silicon wafer, and (c) pipette making direct contact with substrate while printing. (d) Schematic of capillary flow printing of AgNPs. SEM images of two sets of CFP AgNP traces showing (e) well-defined AgNP edges and a uniform gap between the electrodes and (f) an electrically open gap (i.e., transistor channel length) of ~340 nm. (g) Profilometry profile of Duke logo printed with AgNPs. Pictures of ion gel traces printed by (h) CFP and (i) AJP with insets of profilometry showing CFP ion gel's thin trace and defined line

edges (without overspray). (j) Image of 132 devices printed in a 1mm-by-1mm square, demonstrating CFP's ability to scale down device dimensions.

**Submicron printed carbon nanotube transistors**

To demonstrate the utility of CFP in printing CNT-based transistors, we began with a simple back-gated, bottom-contacted, and toluene-rinsed CNT-TFT device on Si/SiO$_2$ (this method uses the fewest steps and follows that used in much of our previous work)[50]. This process requires two printing steps: 1) printing pairs of AgNP electrodes with <1 µm gap between them, followed by sintering of the AgNP patterns at 200 °C for 90 minutes, and 2) printing a CNT ink above the gap, and rinsing the chip in 80 °C toluene for 10 minutes. It is worth noting that completely obviating the use of toluene (both in the ink solvent and in rinsing), which would promote better environmental stewardship, is possible with additional ink optimization[9]. An optical image of the CNTs being printed is shown in Figure 2a, where a visible line is seen due to residual ink solvent (toluene). SEM imaging confirms the presence of a dense and percolated CNT network bridging the submicron channel length (Fig. 2b), thus forming a full CNT-TFT (the schematic of which is shown in Fig. 2c).

Among the most important of the many tunable parameters for CFP are print speed and ink formulation. First, we examined the effect of increasing the stage movement speed during the printing of AgNP electrodes. As shown in Figure 2d, print speed has a significant effect on the quality of the printed line. With a slow print speed (20 µm/s), the line is uniform with low edge roughness, allowing for fine control over the quality of the print, enabling consistency in printing submicron gaps between silver lines. As the speed is increased to 200 µm/s and 2 mm/s, control over the quality of the printed lines decreases, as is evident by the increase in low-density AgNPs at the edges of the printed line. This trend follows with the measured height of the printed AgNPs (Fig. 2e-f), which shows that as the print speed is increased, there is less ink being deposited, meaning that the lines are thinner, but only to a certain point after which the height levels off and the width shrinks instead (Fig. 2g). This verifies the importance of tuning print speed for CFP, depending on the required line fidelity and print density.

Similarly, while there are many types of ink formulation parameters that could vary print quality (e.g., viscosity, volatility, surface tension), CNT ink concentration tends to be the most significant with direct-write printing techniques and was thus our focus in this study. Density of the CNT thin film is critical to control and strongly dependent on ink concentration (Fig. S3); a density that is too low will not yield sufficient percolative transport linkage in the transistor channel while density that is too high will increase the probability of a metallic nanotube electrically shorting the channel. As shown in Figure 2h, the density of the film is relatively low at 25 µg/mL (barely forming a network between the two electrodes), moderate at 37.5 µg/mL (forming a dense cluster in a single thin area), and high at 50 µg/mL (forming a uniform, dense, and consistent film). Thus, 50 µg/mL was chosen as the ink concentration of choice for the remainder of the studies in this work. As a result, it is critical to determine an estimate of the density of CNTs at this ink concentration for accurate $W_{ch}$ and mobility extraction. Software-based density analysis of SEM images of different devices showed a mean density of ~54.3%, with a standard deviation of 4.8% (Fig. S3a-c), and a representative $W_{ch}$ of 15.5 µm (Fig. S4d). The average $W_{ch}$ used for width normalization was 16.4 µm.

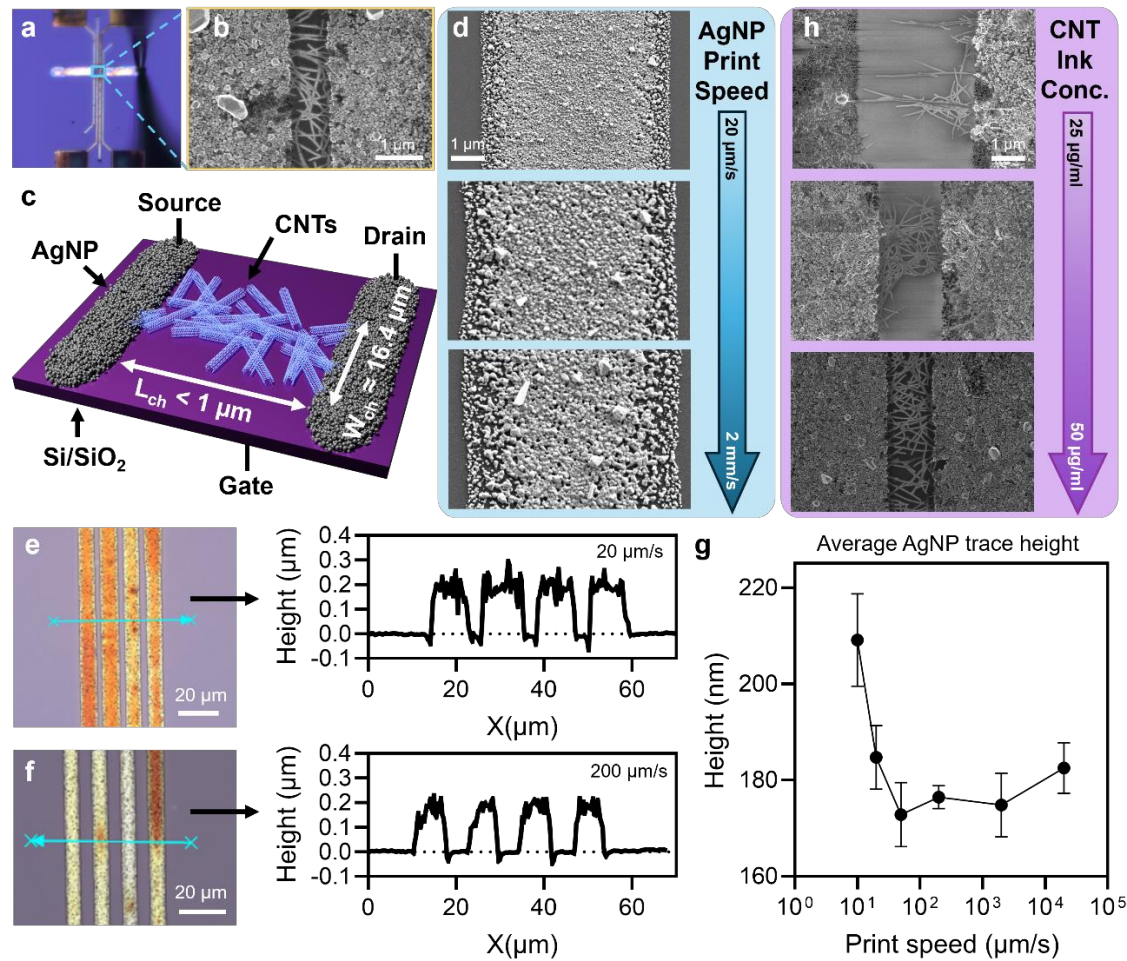

**Figure 2. Capillary flow printing submicron CNT-TFTs.** (a) Image of capillary flow printer depositing CNT ink onto printed AgNP electrodes. (b) SEM image of dense film of CNTs printed onto AgNP electrodes with a sub-µm channel length. (c) Schematic of a capillary flow printed CNT-TFT on 90 nm SiO$_2$ indicating sub-µm channel lengths (L$_{ch}$) can be realized without chemical treatments or modifications. (d) SEM images of capillary flow printed AgNP traces at print speeds of 20 µm/s, 200 µm/s, and 2 mm/s (from top to bottom), realizing denser and more uniform films at slower print speeds – all images at same magnification. Image (left) and profilometry measurement (right) of CFP AgNP electrodes printed at (e) 20 µm/s and (f) 200 µm/s. (g) Plot showing the AgNP trace height change as print speed is varied. (h) SEM images of capillary flow printed CNT traces over AgNP electrodes at CNT ink concentrations of 25 µg/ml, 37.5 µg/ml, and 50 µg/ml (from top to bottom), realizing denser and more uniform films at higher concentrations.

In fabricating the substrate-gated CNT-TFTs, two different contacting architectures were examined: bottom contacts (Fig. 3a) and top contacts (Fig. 3b). Subthreshold (Fig. 3d) and transfer

(Fig. 3e) curves of five representative bottom-contacted devices (all with submicron $L_{ch}$) show the strong device performance of this contact geometry. All five devices showed relatively similar performance, with modest on-currents of ~10-30 µA/mm, but with on/off-current ratios of 3 orders of magnitude. Interestingly, top-contacted devices (Figs. 3e-f) performed much better in terms of uniformity and on-current, reaching values of up to 1000 µA/mm, (average of 956 ± 112 µA/mm) but with a slightly lower average on/off-current ratio. This noticeable improvement in on-current (by ~3 orders of magnitude) for top-contacted CNTs is attributed to the combination of more conformal interfacing of the CNT thin film with the substrate (which improves gating efficiency) as well as better removal of residual wrapping polymer used in the CNT ink by using rapid thermal annealing (RTA) after CNT printing. However, this method is not always favorable as there is a tradeoff between high-temperature baking (a mechanism of obtaining high $I_{on}$) and material compatibility. Toluene-rinsing the chip (soaking in toluene at 80 °C for 10 minutes) is an easier technique that is compatible with all components of the device (Si substrate, Ag electrodes, CNT channel), though it does require the use of an environmentally toxic solvent[9]. Contrarily, RTA avoids that issue and could be even more efficient at removing residual surfactant, but it does require an expensive and relatively energy-intensive tool that requires a higher thermal budget for all materials; in this case, the silver electrodes would corrode under such high heat (500 °C) and lose conductivity. Thus, these findings demonstrate the applicability of CFP with both techniques but emphasize the need to consider the tradeoff between performance, environmental impact, and material considerations. Moreover, these findings are made even more impactful when comparing silver electrodes fabricated using CFP and electron beam lithography/evaporation, which remains the "gold standard" technique for the fabrication of submicron transistors. The extracted metrics for these two types of electrodes show that CFP competes with, and even outperforms cleanroom-based techniques based on these metrics, highlighting its utility as a valuable fabrication technique (Fig. S5).

Post-processing methods like burn-in and sintering could further enhance device performance. Burn-in is a technique where a high electric field is applied across the channel ($V_{DS}$) to induce resistive heating that has a localized sintering effect on the device (Fig. 3g). Figure 3h illustrates the notable improvement in device performance characteristics for a bottom-contacted device; it showed an on-current of 21 µA/mm and an on/off-current ratio of $10^{1.8}$ before burn-in (light blue curve), and an on-current of 145 µA/mm and on/off-current ratio of $10^{3.4}$ after burn-in (dark blue curve). This increase in on-current is a result of burning off the residual insulating polymer, while the decrease in off-current is due to the burning of metallic tubes (note, burn-in was performed while a gate-source voltage was applied to hold the semiconducting CNTs in the off-state). Similarly, longer sinter times also show drastic improvements in the on-current of the device, with up to 672% change in normalized on-current between 20 minutes and 90 minutes of sintering. These two effects illustrate the possibilities of improving device performance through relatively easy and simple techniques.

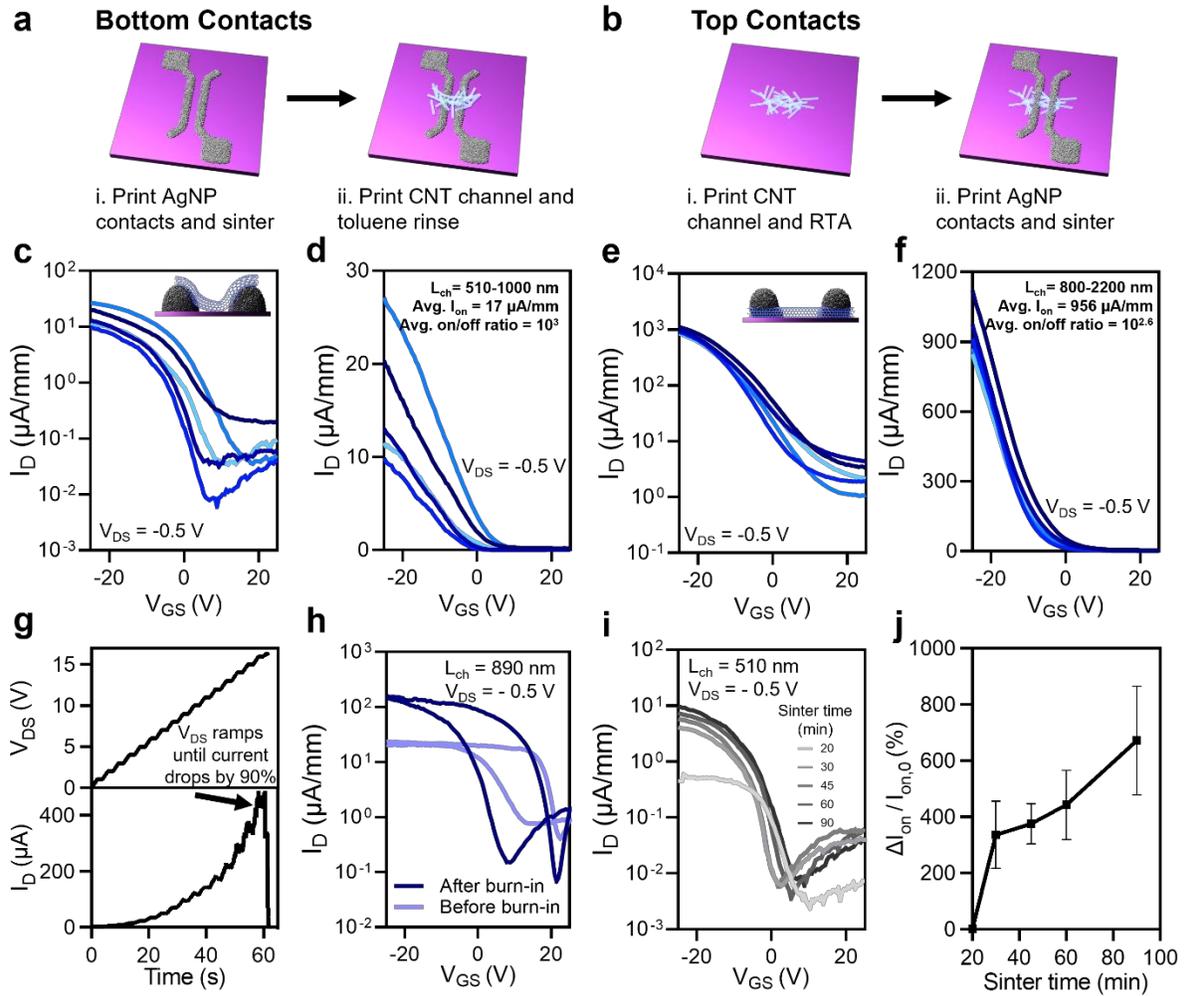

**Figure 3. Electrical characterization of back-gated CFP submicron CNT-TFTs.** Schematic of process flow for printing (a) bottom contact and (b) top contact CNT-TFTs using CFP. Subthreshold (c) and transfer (d) curves for five bottom-contacted CNT-TFTs, with corresponding schematic and extracted parameters inset. Subthreshold (e) and transfer (f) curves for five top-contacted CNT-TFTs, with corresponding schematic and extracted parameters inset. (g) Plot demonstrating bottom-contacted device burn-in over time with $V_{DS}$ ramp up (top) and corresponding $I_D$ response (bottom) under an applied $V_{GS}$ = 15 V for a device that exhibited burnout of metallic CNTs in the thin film. (h) Subthreshold curve for a device before and after burn-in, indicating performance improvement. (i) Subthreshold curves for a bottom contacted device tested after various sintering times at 200 °C. (j) Plot showing an increase in the normalized on-current as the sintering time is increased.

**Fully printed submicron carbon nanotube transistors**

To achieve the full potential of CFP for printed CNT transistors, the devices must be printed onto substrates beyond silicon through the addition of a printed dielectric and gate (Fig. 4a). We printed ion gel as the gate dielectric using AJP or CFP (Fig. 4b-c), modulating the CNT channel with low applied voltages ($V_{DS}$ = -0.5 V, $V_{GS}$ ±1 V) due to the rapid formation of an electric double-layer (EDL) under applied electric fields attributed to the high ionic conductivity of the ion gel[51,52]. Although gating through ion gel degraded the on-, off-, and leakage currents, several device performance characteristics such as hysteresis, SS and threshold voltage were improved along with a reduction in power consumption (Fig. 4c, Fig. S6), highlighting the trade-off between gating through $SiO_2$ and ion gel.

The fully printed submicron CNT-TFTs gated with AJP ion gel (using a well-established ion gel print procedure) were further compared to devices gated with CFP ion gel. The submicron channels were successfully gated through AJP and CFP ion gel of the same formulation, modulating the channel within a $V_{GS}$ of ± 1 V (Fig. 4f-g). When gating through ionic dielectrics, transistor performance depends on the sweep rate of the gate voltage due to the movement of ions within the film that form the EDL[50,51]. In comparing the sweep rate dependence of AJP and CFP ion gel-gated devices, most performance metrics showed similar trends in the measured range between 20 and 430 mV/s (Fig. 4f-g and Fig. S7). Further, at lower sweep rates, the performance characteristics of both ion gels were similar besides a lower hysteresis, leakage current, and subthreshold swing for the CFP ion gel. This study also revealed that the CFP ion gel has a greater resilience to changes in sweep rate for all examined device performance metrics. Notably, a 1550% and 148% change in $I_{off}$ from 20 to 430 mV/s was observed for AJP and CFP ion gel devices, respectively. The robustness and improved performance of the CFP ion gel may be attributed to the thinness of the resultant CFP film (<1 μm) compared to the AJP film (~10–20 μm) (Fig. 1i, 4b), which lowers the resistive-capacitive (RC) delay (i.e. the time it takes to charge the EDL)[53–55]. CFP also produced ion gel traces with cleaner, overspray-free edges, reducing the ion gel coverage over the electrodes and lowering the gate capacitance (Fig. 1h and Fig. 4c).

Due to the superior performance of CFP ion gel, we benchmarked these devices relative to other fully printed CNT-TFTs on flexible substrates shown in literature, comparing channel length and a key performance metric ($I_{on}$). Importantly, at a gate sweep rate of 20 mV/s, the on-current behavior for these fully printed CFP CNT-TFTs (with AJP ion gel) is the highest, especially for a channel length below a micron (Fig. 4g and Fig. S8; data from Supplementary Table 1). The current of the CFP CNT-TFTs rivals that of the IGZO and LTPS devices, at a channel length smaller than a tenth of these PECVD devices. The relatively high on-current at a low drain voltage (490.4 μA/mm at a $V_{DS}$ of -0.5 V) highlights the ability to achieve high-quality films with CFP even on rough surfaces like Kapton. This feat is particularly impressive when considering that these CNT-TFTs simultaneously push the achievable channel length of printed devices to well below a micron and demonstrate the viability of CFP CNT-TFTs for high-density applications on flexible substrates.

To assess the suitability of CFP CNT-TFTs for flexible electronics, submicron channels were fabricated using CFP on Kapton and tested for mechanical robustness (Fig. 4h). Ion gel, a mechanically robust material, was printed using AJP rather than CFP due to the surface roughness of Kapton, which may be overcome with optimization of CFP parameters[50,56]. The resultant 8 devices underwent mechanical stress testing from 0 to 1000 bend cycles around a 2 mm bend radius (Fig. 4i). Interestingly, although all average performance parameters worsened after 1000 bend cycles (besides $I_{off}$), most degradation of the device performance occurred after the initial

100 bend cycles, with some recovery and relative stabilization as bending continued (Fig. S9). This slight change in performance may be attributed to the AgNPs shifting during bending, as the ion gel and CNTs have not been impacted by mechanical bending in previous studies[50,57]. Importantly, the overall performance of the devices remained strong after 1000 bend cycles, demonstrating the mechanical resilience of CFP submicron devices, and paving the way for their use in flexible electronics applications.

Although CFP submicron devices mitigate pollutants by eliminating the use of cleanroom processes, ion gel is toxic to the environment. Therefore, to improve the sustainability of the submicron devices, crystalline nanocellulose (CNC), a biodegradable ionic dielectric[6], was also aerosol jet printed as the gate dielectric. Due to the high ionic resistance of nanocellulose, a recyclable graphene gate was aerosol jet printed over the CNC, covering the CNT channel region[6]. Since CNC is also an ionic dielectric, the gate sweep rate from 20 to 520 mV/s influenced the performance of the device with similar trends to those seen in the ion gel study (Figs. S10). One noteworthy difference was the change in hysteresis direction around 50 mV/s from counterclockwise at slower sweep rates to clockwise at faster sweep rates, attributed to the charging current of the EDL at faster sweep rates (Fig. S10c-d)[9]. The compatibility of CFP submicron transistors with AJP CNC and the ability of CFP to print conductive AgNP electrodes on paper substrates (Fig. S11) are significant steps in improving the sustainability of printed submicron transistors. However, further work must be carried out to create more sustainable CFP transistors such as printing carbon-based electrodes and a water-based CNT ink.

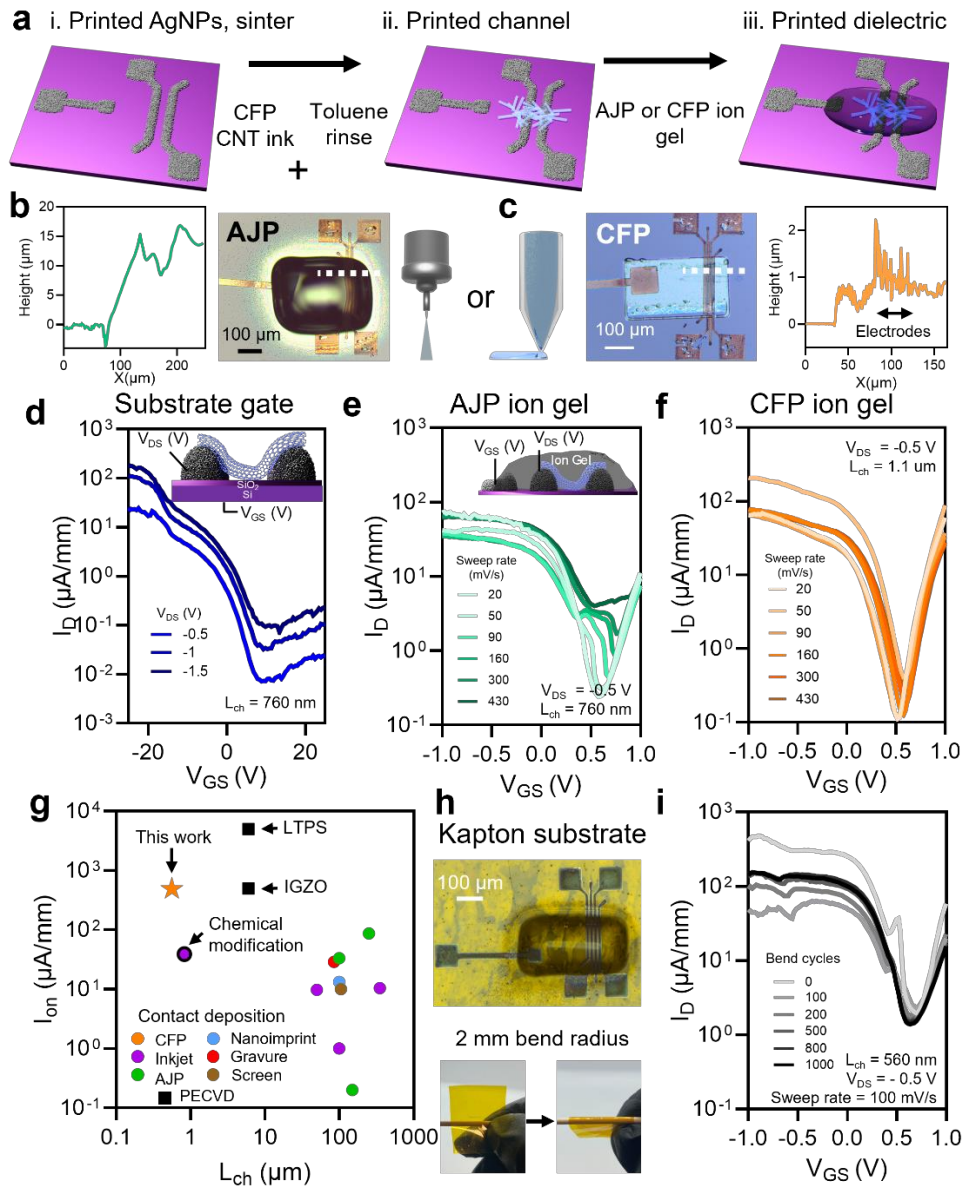

**Figure 4. Fully printed submicron CNT-TFTs.** (a) Printing fabrication schematic for side-gated CNT-TFTs. Note, AgNPs were sintered 90 min at 200 °C, and ion gel gate dielectric was printed over side-gate electrode and channel regions. Profilometry and optical image of (b) AJP and (c) CFP ion gel of the same formulation printed over a device with designed dimensions of 350 μm x 200 μm. (d) Subthreshold curves at different $V_{ds}$ of a device with $L_{ch}$ = 760 nm, substrate gated through $SiO_2$. Subthreshold curves at different sweep rates side gated through (e) AJP ion gel and (f) CFP ion gel on $SiO_2$. (g) Benchmarking plot of width-normalized on-current vs channel length for fully printed CNT-TFTs with Ag source and drain contacts. Data for other CNT-TFTs can be found in Supplementary Table 1. (h) Optical images of a fully printed CNT-TFT with ion gel gate on Kapton, with devices bent around a 2 mm rod to measure mechanical stability. (i) Subthreshold curves of bending cycles from 0 to 1000 cycles around a 2 mm

bending radius for an ion gel-gated device with channel length of 560 nm, revealing minor degradation at 100 bends but remarkable recovery in device performance as bending proceeded. All channel widths ~ 9 μm.

## Conclusions

In this work, we reported the direct printing of submicron channel length CNT-TFTs on both rigid and flexible substrates using a capillary flow printer. Channel lengths as small as 500 nm were repeatedly demonstrated with no chemical modification required using AgNP ink. CFP of CNTs as the semiconducting channel material between submicron-spaced electrodes on $SiO_2$ yielded devices with an on-current of 1.12 mA/mm, indicating the strong performance achievable using this printing technique. The versatility of this technique is emphasized through its use in printing a wide variety of materials (AgNPs, CNTs, SU-8, and ion gel), different contact geometries (bottom and top contacts), and gating architectures (bottom, side, and top gates) on various substrates (Si, Kapton, and paper). Fabrication of CNT-TFTs using an ion gel dielectric and a side gate verified the ability to create fully printed TFTs that outperform other demonstrations of fully printed scaled CNT transistors. Further, the sweep rate and bending resilience of these fully printed devices highlights their utility in flexible electronics applications and beyond. All in all, these findings emphasize the customizable, easy-to-use, and versatile nature of this printing technique for the fabrication of submicron transistors, paving the way toward scalable and cleanroom-free fabrication of scaled electronics.

# Methods

**Materials**

Silver nanoparticle (AgNP) ink was purchased from Hummink with an ink weight fraction of 68 +/- 2 wt% and was printed as purchased. Semiconducting CNTs (IsoSol-S100 polymer-wrapped nanotubes) were purchased from NanoIntegris Inc at a concentration of 50 μg/ml. The as-purchased ink was ultrasonicated for 1 hour to redisperse the CNTs into solution then printed without dilution unless otherwise noted. SU-8 6000.5 photoresist was purchased from Kayaku and printed without dilution. Ion gel was prepared by combining polystyrene-*b*-methyl methacrylate-*b*-styrene (PS(600)-*b*-PMMA(118000)-*b*-PS(600)) purchased from Polymer Source Inc, 1-ethyl-3-methylimidazolium bis(trifluoro-methylsulfonyl)imide (EMIM-TFSI) purchased from Sigma-Aldrich, and ethyl acetate purchased from Sigma-Aldrich in a 1:9:90 ratio by weight. The ion gel was stirred overnight before use in the printers. The crystalline nanocellulose (CNC-Slurry-HS) was purchased from Cellulose Lab at a concentration of 10 wt% solids. To prepare the ink for printing, the CNC was diluted in DI water to a concentration of 6% wt/wt CNC. Sodium chloride (>99.0% ACS reagent grade) was purchased from Sigma-Aldrich and was added to the CNC solution to a concentration of 0.05% wt/wt NaCl. Graphene ink (Sigma-Aldrich at 10 wt% graphene concentration) was diluted with DI water to an approximately 2.33 wt% solution of graphene before printing. Si/SiO$_2$ wafers were purchased from University Wafer Inc with 90 nm of SiO$_2$. The Kapton polyimide film (electrical-grade Kapton polyimide film) was purchased from McMaster-Carr with a 0.001-inch thickness.

**CFP printing**

All printing was performed on a NAZCA printer (Hummink) with a glass pipette diameter of 5 μm. Hummink's autotune was used to find the resonant frequency of each pipette when filled with the specific ink being printed.

*AgNP printing*

All AgNP films on silicon were printed with a travel speed of 20 μm/s, a lift level of 20 μm, and a frequency shift of 500 mHz. On Kapton, AgNP traces were printed with a travel speed of 50 μm/s, a lift level of 20 μm, and a frequency shift of 200 – 500 mHz. One pass of AgNP ink was used in all prints. After printing, the AgNP traces were sintered 200 °C for 90 minutes in an oven.

*Semiconducting CNT printing*

All CNT films on each substrate were printed with a travel speed of 20 μm/s, a lift level of 20 μm, and a frequency shift of 500 mHz. One pass of CNT ink was used in all prints. After printing, the CNT films were either soaked in a toluene bath for 10 min at 80 °C to remove excess wrapping polymer or annealed in a rapid thermal annealing system (Jipelec JetFirst 100) at 500 °C for 8 min with a 2-min temperature ramp to achieve a sufficiently conductive transistor channel.

*SU-8 printing*

All SU-8 6000.5 photoresist films on each substrate were printed with a travel speed of 100 μm/s, a lift level of 20 μm, and a frequency shift of 500 mHz. One pass of SU-8 ink was used in all prints. After printing, the SU-8 film was exposed to UV light (365 nm) for 9 seconds then baked post-exposure for 2 min at 110 °C.

*Ion gel printing*

All ion gel films on each substrate were printed with a travel speed of 50 µm/s, a lift level of 20 µm, and a frequency shift of 500 mHz. One pass of ion gel ink was used in all prints. The ion gel was functional as printed; therefore no post-processing was completed.

**Aerosol jet printing**

All aerosol jet printing was performed on an AJ-300 printer (Optomec).

*Ion gel printing*

A 150 µm diameter nozzle was used to print all ion gel films with the platen temperature at 80 °C and the ink temperature at 20 °C. A print speed of 2 mm/s, a sheath flow rate of 25 SCCM, an atomizer flow rate of 26 – 27 SCCM, and an ultrasonic current of 300–350 mA were used. One pass of ion gel ink was used in all prints.

*Cellulose nanocrystals printing*

A 300 µm diameter nozzle was used to print all CNC films with the platen temperature at room temperature and the ink bath temperature at 20 °C. A print speed of 2 mm s$^{-1}$, a sheath flow rate of 30 SCCM, an atomizer flow rate of 35 SCCM, and an ultrasonic current of 400 – 450 mA were used. One pass of cellulose nanocrystals ink was used in all prints.

*Graphene printing*

A 150 µm nozzle was used to print graphene films with the platen temperature at room temperature and the ink bath temperature at 20 °C. A print speed of 2 mm s$^{-1}$, a sheath flow rate of 25 SCCM, an atomizer flow rate of 37-40 SCCM, and an ultrasonic current of 350 – 400 mA were used. One pass of graphene ink was used in all prints.

**CFP transistor fabrication**

*Back-gated bottom-contacted transistors*

To print the back-gated bottom-contacted transistors, a Si/SiO$_2$ wafer was placed onto the capillary flow printer platen. AgNP source and drain electrodes and contact pads were printed using the above print parameters for the AgNP ink. After AgNP printing, the substrate was baked at 200 °C for 90 min in an oven to achieve conductive AgNP films. The substrate was placed back onto the capillary flow printer platen and the CNT channel was printed using the above parameters. After CNT printing, the substrate was soaked in a toluene bath at 80 °C for 10 min to remove the binding polymer wrapped around the CNTs to achieve a conductive transistor channel.

*Back-gated top-contacted transistors*

To print the back-gated top-contacted transistors, a Si/SiO$_2$ wafer was placed onto the capillary flow printer platen. AgNP alignment marks were printed using the above print parameters for the AgNP ink. Without removing the substrate from the printer, the CNT channel was printed using the above parameters. After CNT printing, the substrate was either soaked in a toluene bath at 80 °C for 10 min or annealed in a rapid thermal annealer at 500 °C under nitrogen for 8 min to remove the binding polymer wrapped around the CNTs to achieve a conductive transistor channel. AgNP source and drain electrodes and contact pads were printed using the above print parameters for the AgNP ink. After AgNP printing, the substrate was baked at 200 °C for 90 min to achieve conductive AgNP films.

*Fully printed ion gel-gated transistors*

To create fully printed transistors, a Si/SiO$_2$ wafer or Kapton film was placed onto the capillary flow printer platen. AgNP source, drain, and side-gate electrodes and contact pads were printed using the above print parameters for the AgNP ink. After AgNP printing, the substrate was baked at 200 °C for 90 min in an oven to achieve conductive AgNP films. The substrate was placed back onto the capillary flow printer platen and the CNT channel was printed using the above parameters. After CNT printing, the substrate was soaked in a toluene bath at 80 °C for 10 min to remove the binding polymer wrapped around the CNTs to achieve a conductive transistor channel. The substrate was placed onto either the aerosol jet printer or Hummink printer and the ion gel gate dielectric was printed (350 x 150 μm square) using the above parameters.

*Fully printed CNC-gated transistors*

The same process was followed as for the fully printed ion gel-gated transistors with the exception of the final two steps where CNC and a graphene top gate were printed using AJP in this case.

**Evaporated contacts transistor fabrication**

*Back-gated bottom- and top-contacted transistors*

First, the contact pads and alignment marks were added to a 90 nm $SiO_2$ substrate. Silicon wafer pieces with 90 nm $SiO_2$ that had been previously cleaned in acetone and IPA were baked at 180 °C for 1 minute to minimize residual solvent before spin coating. The samples were spin-coated with a layer of 950K PMMA A3 followed by a soft bake at 180 °C for 3 min. We used an Elionix ELS-7500 EX electron-beam lithography system using a 50 kV beam with a dose of 700 μC/cm$^2$ to pattern the PMMA, followed by development in a 3:1 IPA/MIBK mixture for 30 seconds before blowing the samples dry with $N_2$. Next, 1 nm Ti and 30 nm Pd was deposited by electron-beam evaporation, followed by metal lift-off in acetone on a hotplate at 80 °C for 5 minutes. Immediately after lift-off, the samples were cleaned in IPA and blown dry. Next, the bottom contacts were added to the substrate using the same procedure as above except 1 nm of Ti and 30 nm of Ag were deposited by electron-beam evaporation to form the metal stack. Then, the CNTs were CFP printed over the deposited bottom contacts and designated top contact regions using the outlined procedure above. Afterwards, the top contacts were added to the substrate once again using the same procedure as above except 1 nm of Ti and 30 nm of Ag were deposited by electron-beam evaporation to form the metal stack.

**CNT density analysis**

CNT films were printed with CFP over AgNP electrodes with the parameters outlined above. A scanning electron microscope (SEM) was used to take images of each print and ImageJ, an image processing software, was used to determine the CNT density within the channel region. After uploading the image to ImageJ, the image threshold was adjusted, setting the image to binary. The brightness and contrast were manually changed to ensure the CNTs were sufficiently distinguishable from the background (Fig. S4a-c). The desired channel region was selected and the area fraction was measured using the built-in ImageJ procedure. The resulting average CNT density was 54.3 ± 4.8%. After calibrating the image dimensions with the scale bar, the width of the channel was also measured (Fig. S4d). The resulting average width of the channel was 16.4 ± 1.2 μm.

**Instrumentation and characterization**

SEM (ThermoFisher Scientific Apreo S) images and profilometry (Keyence VK-X3050) measurements were taken at the Shared Materials Instrumentation Facility (SMIF) at Duke. All electrical TFT measurements were completed with a manual analytical probe station connected to SMUs (Keysight B2902A).

**Device parameter extraction**

All extracted data was on the forward sweep (-$V_{GS}$ to +$V_{GS}$) other than hysteresis. The on-current was taken as the maximum current of the device in the p-type (largest magnitude negative gate voltage) regime. The off-current was taken as the minimum current of the device. The subthreshold swing was calculated as the minimum inverse slope of $I_D$ on the p-type branch of the subthreshold curve averaged over ~0.2 V from maximum to minimum current. The gate leakage current was the average measured gate current at gate voltages at 20% of the maximum $V_{GS}$ for the forward and backward sweep. The threshold voltage was taken at 0.1 µA for the fully printed devices. Since the performance of the substrate-gated devices is variable, the threshold voltage ($V_{th}$) was taken at an $I_D$ value halfway between the minimum and maximum current. The hysteresis was measured as the voltage difference between the forward and reverse sweep for that same $I_D$ value. The transconductance was the slope of the best fit line between 85% and 70% of the maximum $V_{GS}$ on the p-branch for all devices.

# Associated Content

**Supporting Information**

Supporting information can be found free of charge on the Nature Publications website.

*List of figures:*
Figure S1. CAD file schematic and annotated capillary flow printer components.
Figure S2. CFP SU-8 for biosensors.
Figure S3. Effect of CNT ink concentration on CFP CNT-TFTs.
Figure S4. CNT density analysis.
Figure S5. Comparison of extracted performance metrics for CFP and evaporated Ag electrodes in submicron CNT-TFTs.
Figure S6. Comparison of CFP CNT-TFTs gated through SiO2 back gate and AJP ion gel side gate.
Figure S7. Sweep rate dependence of CFP and AJP ion gel-gated CNT-TFTs.

**Author Information**

**Corresponding Author**

* Email: aaron.franklin@duke.edu; Tel: (919) 681-9471

**Author Contributions**

B.N.S., F.M.A, and A.D.F. conceived the study. B.N.S., F.M.A., J.L.D, X. P., Q. M., M. S., and D. B. fabricated and tested the devices. B.N.S. and F.M.A. contributed to figure design and data analysis. M.P., P.B., M.B., and A.M. contributed to AgNP ink development and print conditions for submicron features. A.D.F. provided scientific guidance and supervised the research project. B.N.S, F.M.A. and A.D.F. wrote the manuscript with revision and approval from all authors.

**Ethic declarations**

M.P., P.B., M.B., and A.M. are employees of Hummink, which develops and sells the NAZCA capillary flow printer.

# Acknowledgements

We gratefully acknowledge support from the National Institutes of Health (NIH) under award no. 1R01HL146849 and from the Pratt School of Engineering at Duke University through the Beyond the Horizon program. B.N.S. acknowledges support from the National Science Foundation Graduate Research Fellowship under Grant No. 2139754. This work was performed in part at the Duke University Shared Materials Instrumentation Facility (SMIF), which is a member of the North Carolina Research Triangle

Nanotechnology Network (RTNN), and is supported by the National Science Foundation (Grant ECCS-1542015) as part of the National Nanotechnology Coordinated Infrastructure (NNCI). The content presented in this manuscript represents the views of the authors and does not necessarily represent the views of the funding organizations.

# Capillary Flow Printing of Submicron Carbon Nanotube Transistors

Brittany N. Smith[a,§], Faris M. Albarghouthi[a,§], James L. Doherty[a], Xuancheng Pei[a], Quentin Macfarlane[a], Matthew Salfity[a], Daniel Badia[a], Marc Pascual[c], Pascal Boncenne[c], Nathan Bigan[c], Amin M'Barki[c], and Aaron D. Franklin[a,b]

[a] Electrical and Computer Engineering Department, Duke University, Durham, NC 27708, USA

[b] Chemistry Department, Duke University, Durham, NC 27708, USA

[c] Hummink Inc, 75003 Paris, France

[*] Correspondence to: aaron.franklin@duke.edu, TEL: +1-919-681-9471

[§] B.N.S. and F.M.A. contributed equally to this paper

**List of figures:**

**List of tables:**

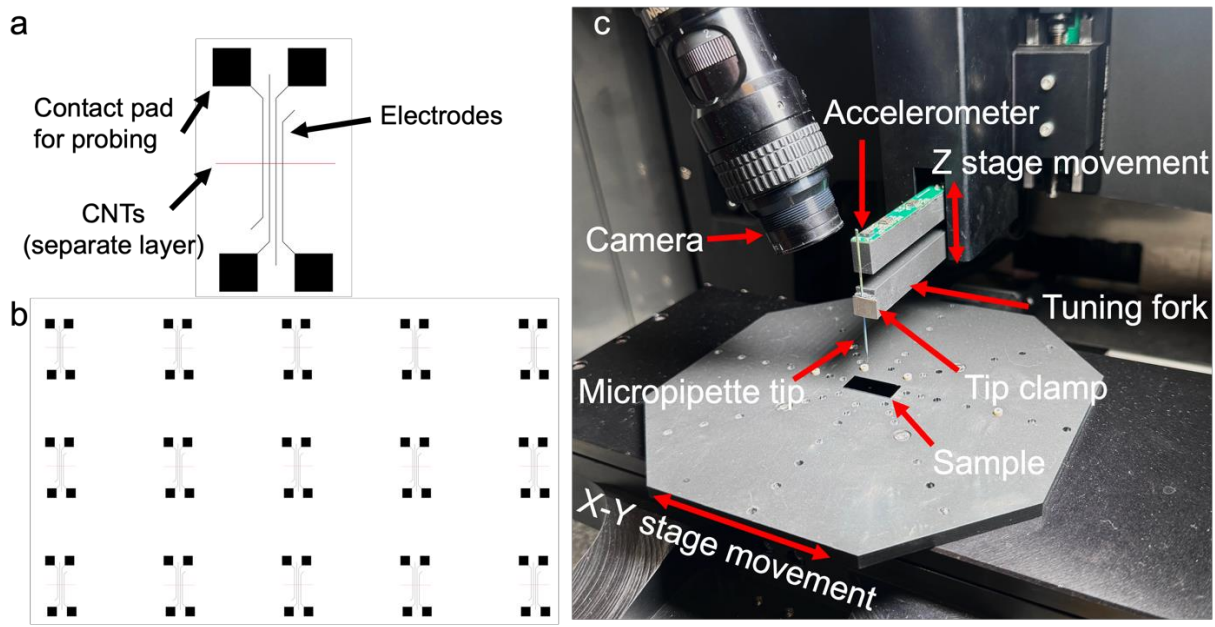

**Figure S1. CAD file schematic and annotated capillary flow printer components.** (a) CAD schematic of a single electrode set containing 4 contact pads for probing with corresponding electrodes, and a CNT layer printed separately with a different ink. (b) A 3x5 array of these electrode sets which are typically printed in a single print run on a chip. (c) Annotated image of key components on the NAZCA capillary flow printer.

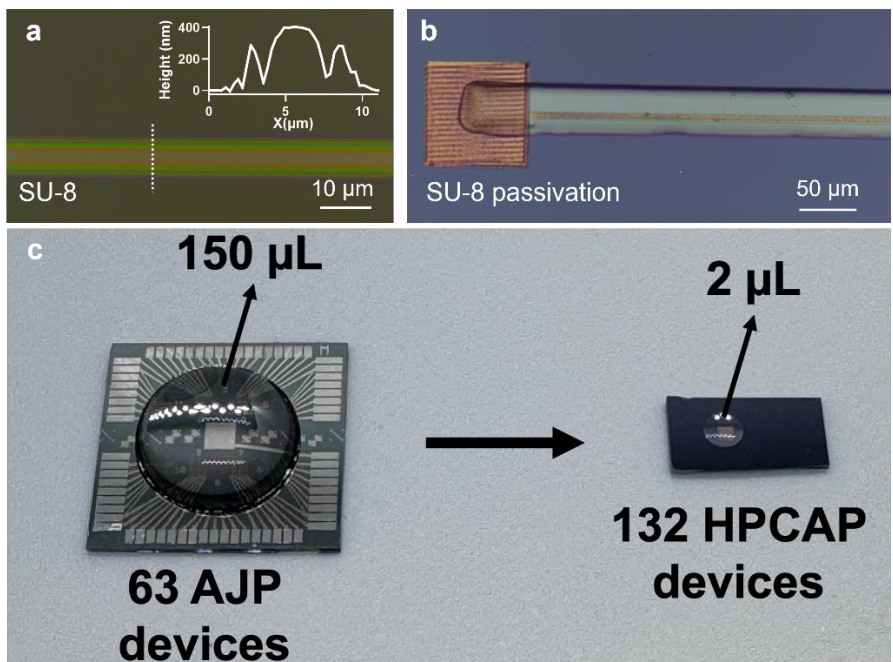

**Figure S2. CFP SU-8 for biosensors.** (a) Picture of SU-8 trace as printed by the CFP with inset of profilometry showing a sub-400 nm thick trace. (b) Picture of lithography-free passivation using SU-8 printed over an AgNP metal electrode. (c) CFP of AgNPs, CNTs, and SU-8 allows for miniaturization of printed biosensors, simultaneously reducing required fluid size from the user and increasing number of devices.

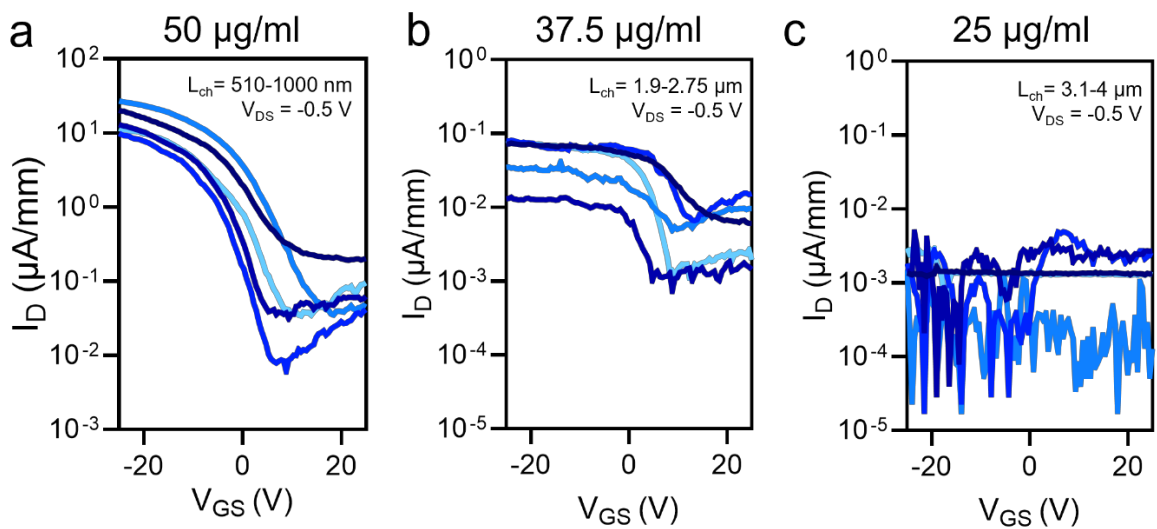

**Figure S3. Effect of CNT ink concentration on CFP CNT-TFTs.** Subthreshold curves for five bottom-contacted CNT-TFTs substrate gated through $SiO_2$ with CNT ink concentrations of (a) 50 µg/ml, (b) 37.5 µg/ml, and (c) 25 µg/ml printed at 50 µm/s and rinsed in toluene, showing the importance of higher CNT concentrations.

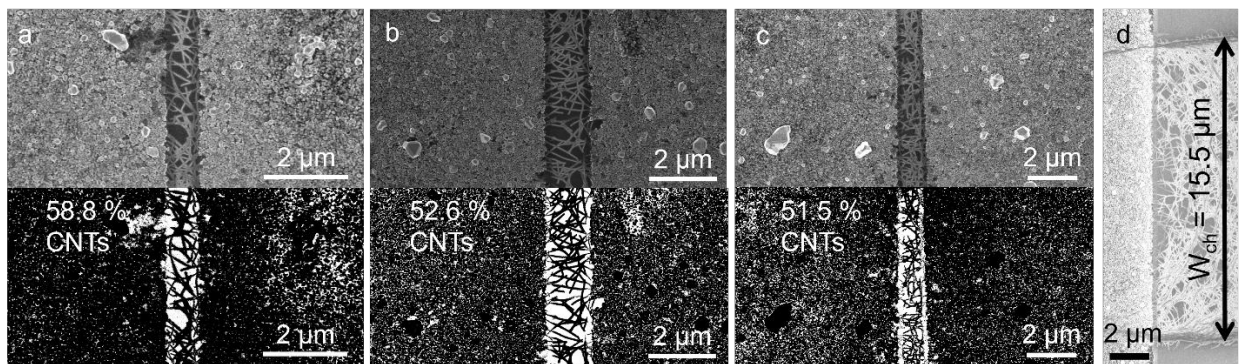

**Figure S4. CNT density analysis.** (a-c) SEM images of CNT thin films printed on silicon. Bottom images set to binary using ImageJ to determine printed CNT density, averaging 54% CNT surface coverage in channel region. (d) SEM image of CNT thin film printed on silicon showing a $W_{ch}$ = 15.5 µm.

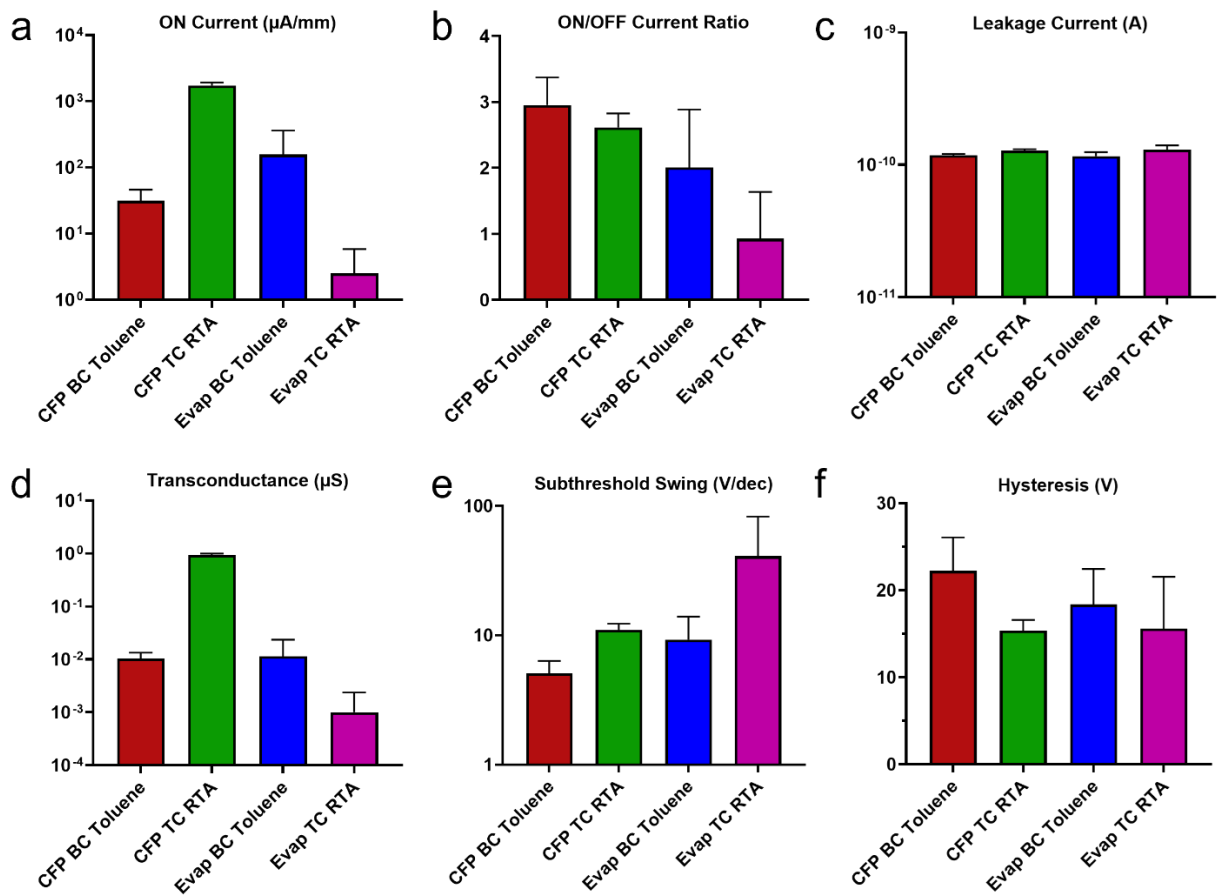

**Figure S5. Comparison of extracted performance metrics for CFP and evaporated Ag electrodes in submicron CNT-TFTs.** Extracted (a) on-current, (b) on/off-current ratio, (c) leakage current, (d) transconductance, (e) subthreshold swing, and (f) hysteresis. For each plot, BC stands for bottom contact, TC stands for top contact, RTA stands for thermal anneal, and Evap stands for evaporated contacts.

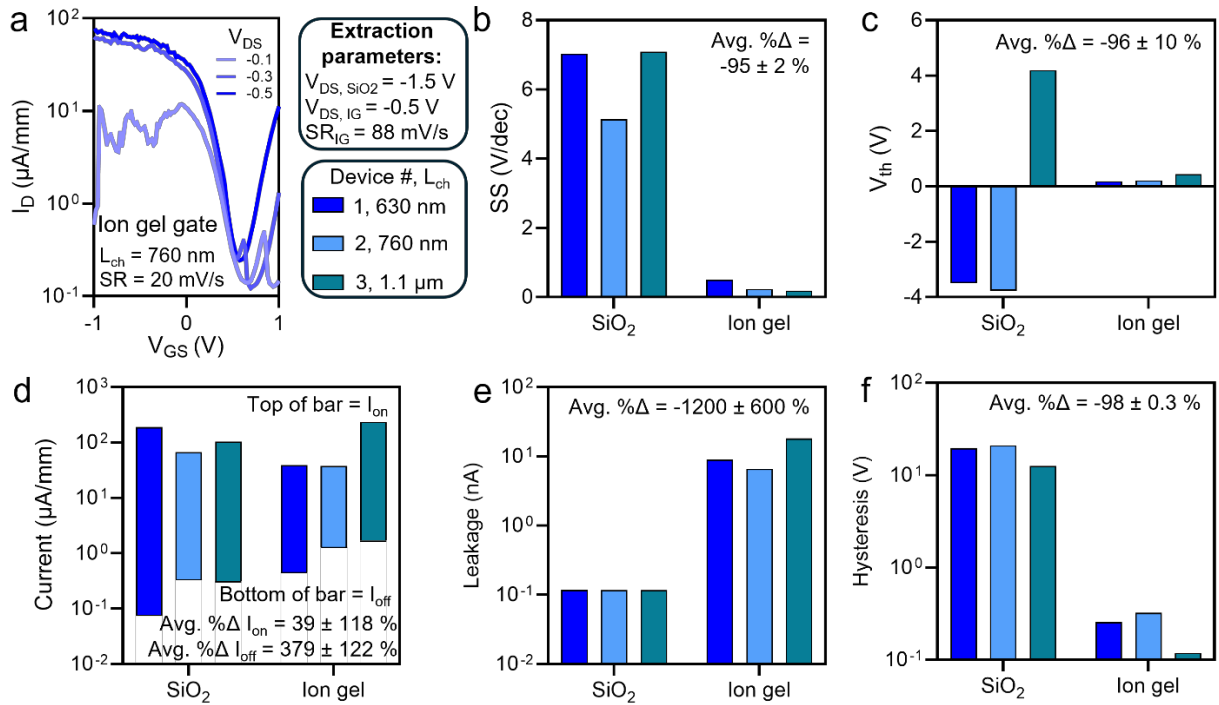

**Figure S6. Comparison of CFP CNT-TFTs gated through $SiO_2$ back gate and AJP ion gel side gate.** Note, the same devices are measured using either of the gates with the back gate ($SiO_2$) below and the ion gel gate above the CNT channel. (a) Subthreshold curves at $V_{DS}$ of -0.1, -0.3, and -0.5 V at 20 mV/s. using ion gel gate. Extraction parameters and $L_{ch}$ outlined for devices gated through $SiO_2$ and ion gel, including (b) SS, (c) $V_{th}$, (d) $I_{on}$ and $I_{off}$, (e) gate leakage, and (f) hysteresis for $SiO_2$ and ion gel-gated devices.

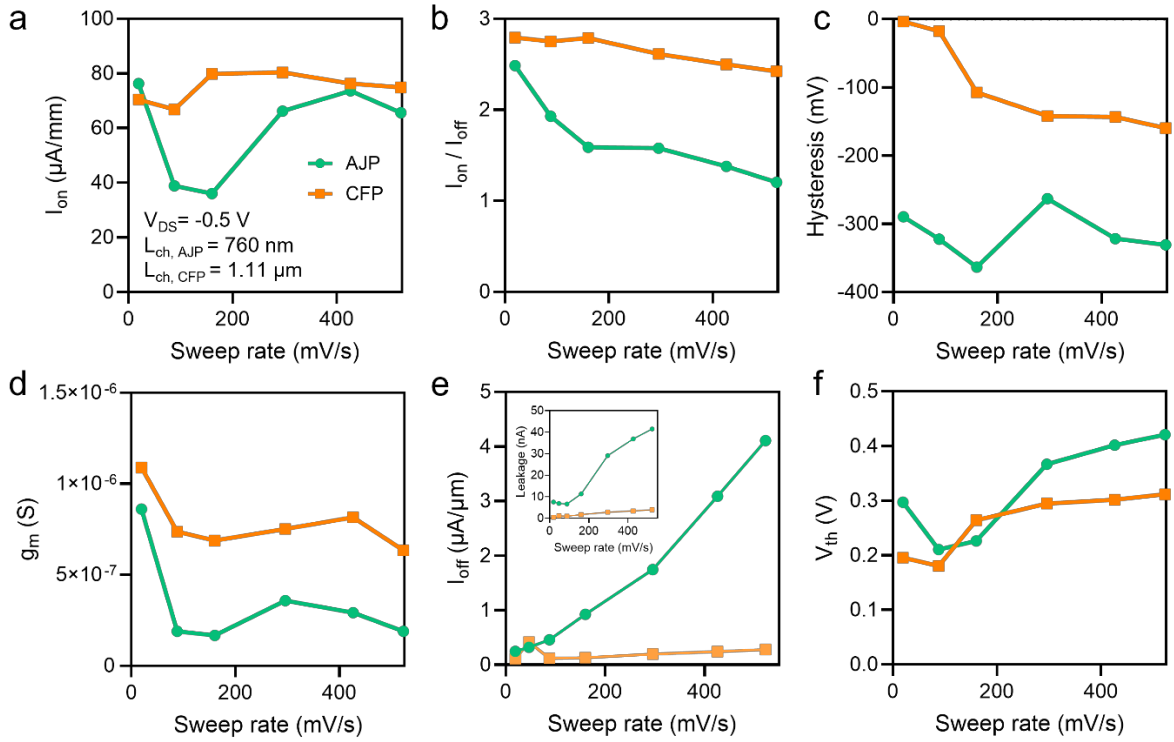

**Figure S7. Sweep rate dependence of CFP and AJP ion gel-gated CNT-TFTs.** Sweep rate dependence of (a) $I_{on}$, (b) $I_{on}/I_{off}$, (c) hysteresis, (d) $g_m$, (e) $I_{off}$ (leakage inset), and (f) $V_{th}$, revealing CFP ion gel is more resilient to changes in sweep rate.

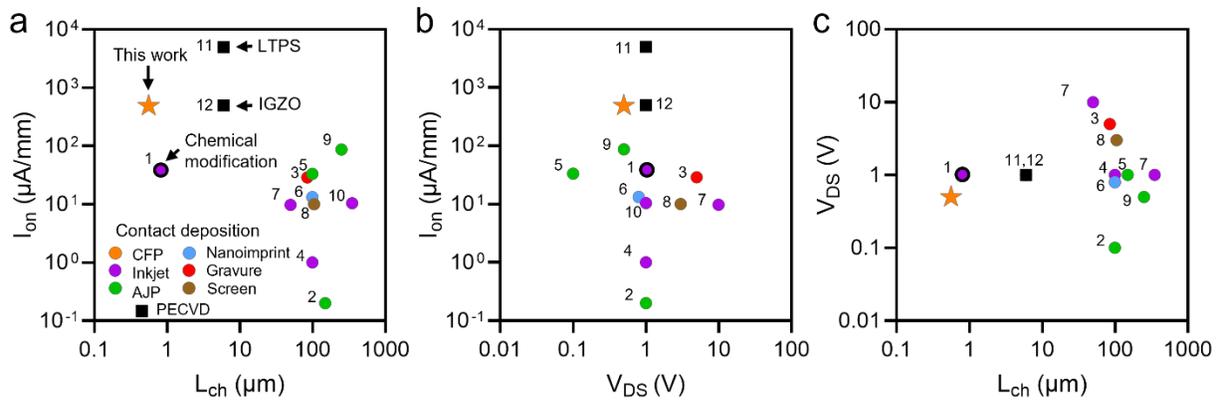

**Figure S8. Extended benchmarking data for fully printed CNT-TFTs with Ag source and drain contacts.** Benchmarking plots of (a) width-normalized on-current vs channel length, (b) width-normalized on-current vs $V_{DS}$, and (c) $V_{DS}$ vs $L_{ch}$.

**Table S1. Comparison table between this work and previous reports on fully printed CNT-TFTs.** PI stands for polyimide, IG stands for ion gel, and (*) marks values that were estimated from graphs in the reports.

| Ref | Substrate material | Electrode material | Dielectric | Electrode fabrication | $L_{ch}$ (μm) | W (μm) | $V_{GS}$ (V) | $V_{DS}$ (V) | $I_{on}$ (μA/mm) |
|---|---|---|---|---|---|---|---|---|---|
| **This work** | **PI** | **AgNP** | **IG** | **CFP** | **0.6** | 16.4 | **1** | **0.5** | **490.4** |
| 1 | PI | AgNP | AZ5214 | Inkjet | 0.8 | 12.5 | 30 | 1 | 40* |
| 2 | PI | AgNP | xdi-dcs | AJP | 150 | 500 | 6 | 1 | 0.2* |
| 3 | PET | AgNP | BTO/PMMA | Gravure | 85 | 1250 | 10 | 5 | 29 |
| 4 | PI | AgNP | BaTiO3/PMMA | Inkjet | 100 | 200 | 10 | 1 | 1* |
| 5 | PI | AgNW | IG | AJP | 100 | 300 | 1 | 0.1 | 33.3* |
| 6 | PET | Ag | IG | Nanoimprint | 100 | 300 | 1.5 | 0.8 | 13.3* |
| 7 | PI | AgNP | BTO/PMMA | Inkjet | 50 | 2000 | 40 | 10 | 9.8 |
| 8 | PET | Ag | BTO | Screen | 105 | 1000 | 10 | 3 | 10* |
| 9 | Photo paper | Graphene | CNC | AJP | 250 | 200 | 2 | 0.5 | 87 |
| 10 | PEN | Ag | PVP | Inkjet | 350 | 300 | 15 | 1 | 10.4* |
| 11 | Si (LTPS) | Mo | $SiO_2$ | PECVD | 6 | 20 | 15 | 1 | 5000* |
| 11 | Si (IGZO) | Mo | $SiO_2$ | PECVD | 6 | 20 | 15 | 1 | 500* |

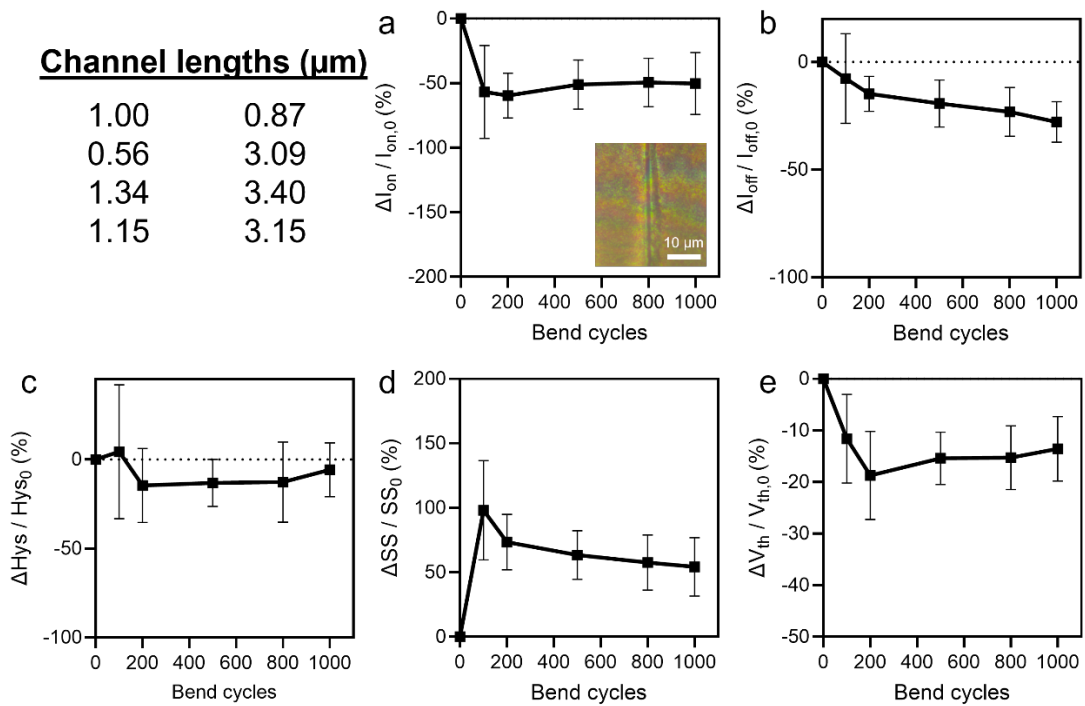

**Figure S9. Bending resilience of CFP CNT-TFTs around a 2 mm bend radius.** Percent change of performance parameters over 1000 bend cycles with respect to each performance metric before bending the device. Performance parameters include (a) $I_{on}$ with an inset image of the 870 nm gap on kapton, (b) $I_{off}$, (c) hysteresis (hys), (d) SS, and (e) $V_{th}$. Error bars encompass 8 devices ranging in $L_{ch}$ from 560 nm to 3.4 μm.

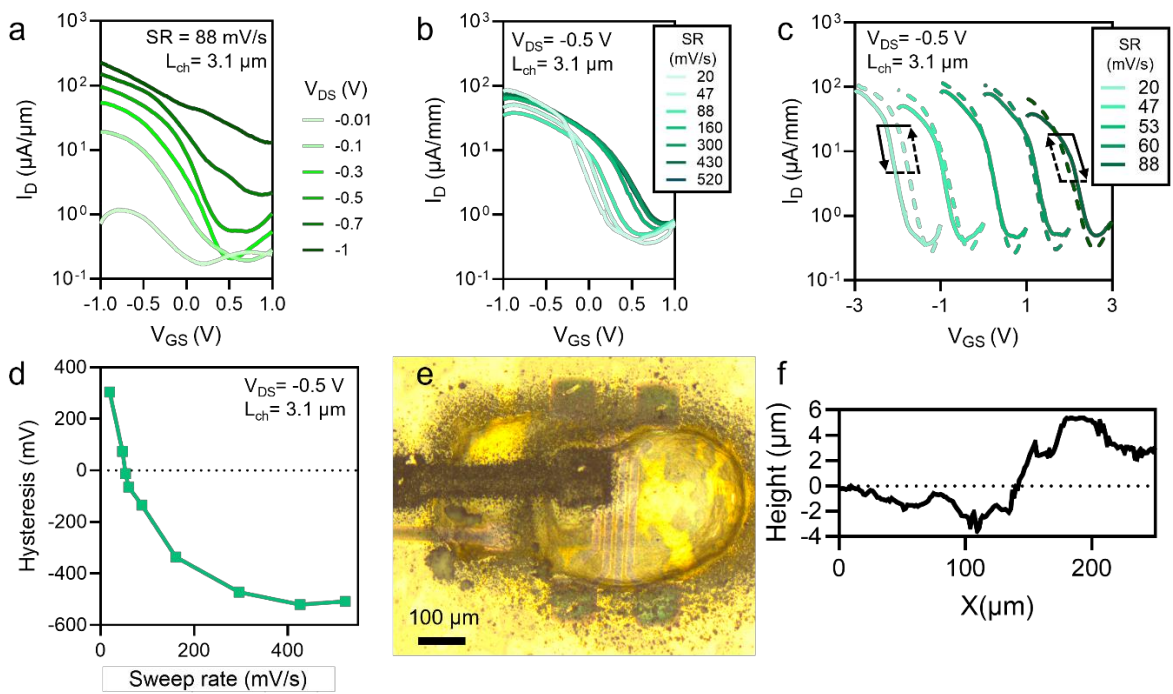

**Figure S10. Top gating CNT-TFTs through CNC.** (a) Subthreshold curves at different $V_{DS}$ revealing the optimal $V_{DS}$ for CNC is -0.5 V. (b) Subthreshold characteristics depending on gate voltage sweep rate. (c) Hysteresis characteristics depending on gate voltage sweep rate. The subthreshold curves corresponding to 20 mV/s, 47 mV/s, 60 mV/s, and 88 mV/s sweep rates are shifted by -2 V, -1 V, 1 V, and 2 V, respectively, to make variations between sweeps visible. (d) Extracted hysteresis plotted as a function of sweep rate. (e) Image of top-gated CNT-TFT through CNC, where source, drain, and channel were CFP printed and CNC and top gate were AJP printed for all CNC top-gated devices. (f) Profilometry of AJP printed CNC film.

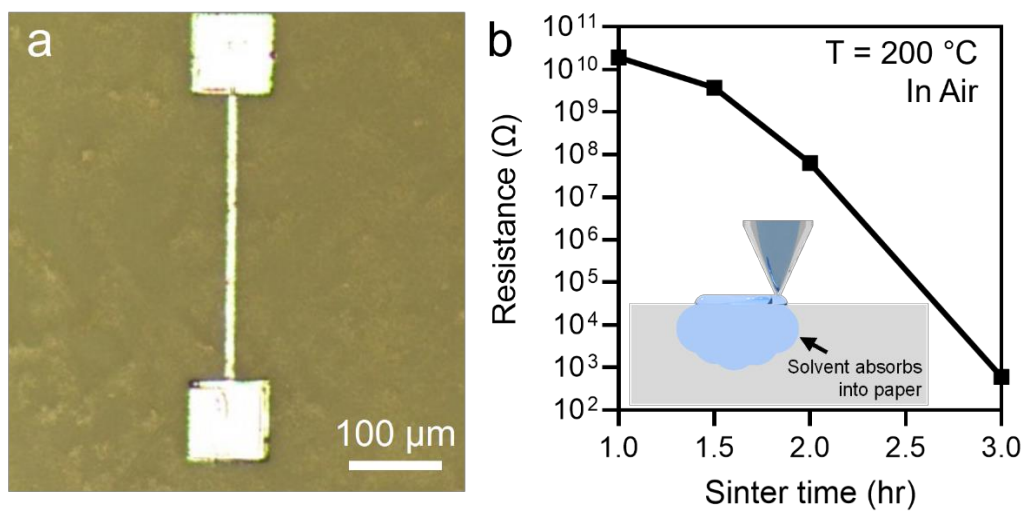

**Figure S11. CFP conductive traces on paper.** (a) Image of conductivity test structure used for resistance measurements of a single line pass of CFP AgNPs. (b) Resistance plot of CFP AgNPs after sintering for 1 to 3 hours, showing a resistance < 1 kΩ. Inset of CFP on paper, illustrating absorption of solvent into paper.